# Topological Protection of Photonic Path Entanglement

Mikael C. Rechtsman[1], Yaakov Lumer[2], Yonatan Plotnik[2], Armando Perez-Leija[3], Alexander Szameit[3], and Mordechai Segev[2]

[1]*Department of Physics, The Pennsylvania State University, University Park, PA 16802*

[2]*Department of Physics, Technion – Israel Institute of Technology, Haifa 32000, Israel*

[3]*Institute of Applied Physics, Friedrich-Schiller-Universität, Jena, Germany*

The recent advent of photonic topological insulators has opened the door to using the robustness of topologically protected transport - originated in the domain of condensed matter physics – in optical devices and in quantum simulation. Concurrently, quantum walks in photonic networks have been shown to yield exponential speedup for certain algorithms, such as Boson sampling. Here we theoretically demonstrate that photonic topological insulators can robustly protect the transport of quantum information through photonic networks, despite the presence of disorder.

Topological insulators are materials that have a bulk band gap, but have edge (or surface) states with energies that cross the gap [1]. In 2D, these states are protected from scattering: they cannot scatter into the bulk due to the gap, and they cannot scatter backwards because the backwards channel is not present or is forbidden from coupling. This leads to robust transport, offering potential applications in quantum information [1], wherein qubits are protected against decoherence.

Photonic topological insulators (PTIs) were proposed [2,3] and realized in microwaves [4]. Then, other schemes were predicted for the optical regime [5–8], and realized in experiments using arrays of helical waveguides [9] and in resonator arrays [10]. In [9], classical light in the paraxial regime diffracted through a honeycomb lattice of helical waveguides, which is mathematically equivalent to the Schrodinger equation. There, a band gap exists in the spatial spectrum, with topological edge states. Each of the bands acquires non-zero Chern number. The experiment showed that even in the presence of scattering (by defects and corners), the edge wavefunction propagates unimpeded.

Independently, quantum walks [11] in photonic lattices have shown rich physics [12–15]. A breakthrough came in 2013, when it was shown that non-interacting quantum walks give exponential speedup in calculating hard-to-compute quantities [16] – this is "Boson sampling". This, taken together with the "KLM protocol" [17] show that non-interacting optical systems show great potential for quantum information processing. As of yet, however, there has been no notion on how topological photonic systems can "topologically protect" quantum information. Indeed, there is no need to protect photons from decoherence because photons barely interact and decohere slowly. So what does it mean to protect *photonic* quantum information?

Here, we show that PTIs can be used to robustly transport fragile multi-photon states in quantum walks. We show that these states maintain their path entanglement despite disorder, in stark contrast with non-topological systems.

First, we show using a simplified analytical argument that scattering will necessarily destroy maximally spatially entangled states (for example, two-particle NOON states [18,19], where the measurement of a photon in one channel implies the other will be observed in the same one). Specifically, we show that upon backscattering, any initially

NOON-state wavepacket will necessary contain non-zero amplitude to measure one transmitted and one reflected photon, making the wavefunction non-NOON.

Consider the diagram of the 1D non-topological lattice with a defect (Fig. 1). The two single-photon spatial states in which the photons are initially launched are $I_l^\dagger|0\rangle$ and $I_r^\dagger|0\rangle$ (denoting left and right wavepackets). We allow the wavepackets to hit the defect and be reflected and transmitted. Assuming that both the left and right wavepackets get reflected and transmitted with the amplitudes $r$ and $t$, we have:

$$I_l^\dagger \to \mathcal{U} I_l^\dagger \mathcal{U}^\dagger = rR_r^\dagger + tT_l^\dagger \text{ and } I_r^\dagger \to \mathcal{U} I_r^\dagger \mathcal{U}^\dagger = rR_l^\dagger + tT_r^\dagger, \quad (1)$$

where $\mathcal{U}$ denotes the evolution operator corresponding to the field Hamiltonian (not the single-particle evolution operator in the Schrödinger picture); $R_l^\dagger$ ($T_l^\dagger$) and $R_r^\dagger$ ($T_r^\dagger$) are the left and right reflected (transmitted) wavepackets, respectively. Consider an initial NOON state: $|\psi(0)\rangle = \frac{1}{2}(I_l^\dagger I_l^\dagger + I_r^\dagger I_r^\dagger)|0\rangle$.

After the state evolves and gets reflected from and transmitted by the defect, the wavefunction is:

$$|\psi(0)\rangle \to |\psi(t)\rangle = \frac{1}{2}\mathcal{U}(I_l^\dagger I_l^\dagger + I_r^\dagger I_r^\dagger)\mathcal{U}^\dagger|0\rangle = \frac{1}{2}[(\mathcal{U} I_l^\dagger \mathcal{U}^\dagger)(\mathcal{U} I_l^\dagger \mathcal{U}^\dagger) + (\mathcal{U} I_r^\dagger \mathcal{U}^\dagger)(\mathcal{U} I_r^\dagger \mathcal{U}^\dagger)]|0\rangle.$$

$$= \frac{1}{2}[(rR_r^\dagger + tT_l^\dagger)(rR_r^\dagger + tT_l^\dagger) + (rR_l^\dagger + tT_r^\dagger)(rR_l^\dagger + tT_r^\dagger)]|0\rangle$$

$$= \frac{1}{2}[r^2 R_r^\dagger R_r^\dagger + t^2 T_l^\dagger T_l^\dagger + 2tr T_l^\dagger R_r^\dagger + r^2 R_l^\dagger R_l^\dagger + t^2 T_r^\dagger T_r^\dagger + 2tr T_r^\dagger R_l^\dagger]|0\rangle. \quad (2)$$

This state is fundamentally non-NOON, because of the amplitudes that mix the transmitted state on the right and the reflected state on the left (and vice versa) – these are the cross-terms. In a fully disordered system, the state will lose more NOONity with each scattering event, hence the final state is necessarily some random state that is not maximally path-entangled. This simplified model also conveys why NOONity is not destroyed in topologically protected lattices. If *r=0* (a property of topological protection), clearly there are no cross terms and the state remains NOON.

To study a quantum walk in a PTI lattice, consider a honeycomb lattice of helical waveguides akin to the design in Ref. [9], and as depicted in Fig. 2(a), i.e., a PTI. The dynamics of the diffraction of a photon through a PFTI waveguide array is given by [9]:

$$i\partial_z a_n^\dagger = \sum_{\langle m \rangle} t e^{iA_0(\cos\Omega z, \sin\Omega z) \cdot r_{mn}} a_m^\dagger + u_n a_n^\dagger \equiv \sum_m H_{nm}(z) a_m^\dagger, \qquad (3)$$

where $z$ is the distance of propagation along the waveguide axis; $a_n^\dagger$ creates a photon on waveguide $n$; $t$ is the coupling; $A_0 = kR\Omega a$ is the gauge field strength (arising due to the helicity); $k$ is the wavenumber, $R$ is the helix radius, $\Omega$ is the spatial frequency of the helices and a is the lattice constant; $u_n$ is a random number lying in the range [-W,W], representing disorder (random waveguide refractive index); and $H^F(z)$ is the $z$-dependent Schrödinger-picture Hamiltonian. Here, $z$ takes the place of time in the usual Schrödinger equation. Thus, the photon diffraction maps to the temporal motion of a quantum particle. This mathematical equivalence between Eq. (3) and the Schrödinger equation has been exploited to probe a wealth of phenomena, including Bloch oscillations [20,21], Zener tunneling [22], Shockley states [23], bound states in the continuum [24], Anderson localization [25–27], photonic quasicrystals [28], photonic graphene [29], and others.

It has been shown [30] that Floquet topological insulators in the strong-driving limit (helix pitch smaller than coupling length), the $z$-dependence can be removed and the system can be described by the Haldane model [31]. We work exclusively in this limit. The practical difference between the two lies in the fact that there is some bending loss associated with the waveguide helicity. Experimentally, this will manifest in a lower photon count, meaning a longer integration time is required. But this does affect photon correlations in the lattice.

This photonic system exhibits topological edge states residing in the bulk gap [9,31] – see Fig. 2(b). Moreover, there are no counterpropagating states in the gap, meaning that when the edge states encounter a defect, they do not scatter. We label the nearest-neighbor coupling term $t_1$, the second-neighbor term $|t_2|=0.2|t_1|$; the coupling phase is set to $\pi/2$ [31]. This regime is easily accessed in the model in Ref. [9]. The band gap is called topological because there is a non-zero topological invariant (the Chern number – see Ref. [31]). The edge states travel to the right along the upper edge and to the left along the bottom edge. Figure 2(c) depicts the structure of the lattice: states are injected on the "clean" left side (i.e., not disordered – W=0), propagate to the right and enter a disordered region. The goal here is to examine the effect of the disordered region on the properties of a two-photon wavefunction. We note in addition that here, we restrict our analysis to pure quantum states being launched into the waveguide array, rather than mixed states. That said, the correlation map associated with ideal NOON states cannot arise from classical light of any form (see criteria for classical light set forth in Ref. [13], which NOON states violate).

Photonic quantum walks involve injecting a photon into a waveguide, or multiple photons in path entangled sets of waveguides [12–15]. The initial state is a superposition of topological edge states. In particular, consider the projection of a single waveguide excitation on only edge modes in the band gap (which can be achieved experimentally through the use of a spatial light modulator to launch the exact wavefunction). We construct these states from the edge states in the band gap within some finite bandwidth, centered in the middle of the gap such that $|E|<E_b$ (where $E$ represents the energy of the state, and $2E_b$ is the bandwidth). A large bandwidth means the edge wavefunction has a small spatial extent, whereas a small bandwidth means a larger extent along the edge. We call $w_n^\dagger$ the operator that creates a photon in the state centered on waveguide $n$. These wavefunctions, which are localized to the top edge of the lattice, are depicted by the red ellipses in Fig. 2(c). They propagate to the right and enter the disordered region (depicted in Fig. 2(c)). Since the edge states only occupy a fraction of the complete spectrum, the wavepackets are 'sinc-like' – i.e., they have decaying outer lobes. Despite the disorder, since the waveguide array acts as a completely closed system obeying deterministic dynamics, there is no mechanism that can lead to loss of phase coherence (i.e., there is no external bath). Thus a multi-photon wavefunction entering the array in a pure state remains in one.

Now, consider the injection of two path-entangled photons along the edge. The initial wavefunction, which contains the amplitude to observe a photon at waveguide $m$ and $n$ can be written $|\psi(z=0)\rangle = \sum_{mn} c_{mn} a_m^\dagger a_n^\dagger |0\rangle$. The correlation map is given by $\Gamma_{mn}(z) = \langle\psi(z)|a_m^\dagger a_n^\dagger a_n a_m|\psi(z)\rangle$ [15] for the two-photon wavefunction $|\psi(z)\rangle$ at propagation distance $z$. As we show in the EPAPS section, $\Gamma_{mn}$ at any propagation distance $z$ can be written in terms of the one-photon propagator, $U(z) = e^{-iHz}$, as $\Gamma_{mn} = |(U(c+c^T)U^T)_{mn}|^2$. Although this expression is general, henceforth we use $m$ and $n$ to index waveguides along

*only the edge* – not the bulk. The expression $P_{mn} = \Gamma_{mn}/(1 + \delta_{mn})$ gives the probability of observing one photon in waveguide *m* and another photon in *n*.

We study the dynamics of two distinct two-photon initial states: a NOON state [18,19], namely $|\psi_{NOON}\rangle = (w_i^\dagger w_i^\dagger + w_{i+l}^\dagger w_{i+l}^\dagger)|0\rangle/2$ ; as well as an "anti-NOON" but still indistinguishable photon state: $|\psi_{SEP}\rangle = (w_i^\dagger w_{i+l}^\dagger)|0\rangle$. Here, *l* indexes the number of waveguides between the states; we take *l=16*. These states may be constructed experimentally using parametric down-conversion and beam shaping with a spatial light modulator.

Figure 3 shows the dynamics of the two-photon states in two cases: for a non-disordered topological system (first row), and for a similar system, but with the right section highly disordered (second row) – corresponding to Fig. 2(c). For all cases, the disorder strength is *W=t/2*. We choose this value of the disorder because it is large enough to cause significant scattering in the trivial case, but not large enough to close the topological gap, negating the topological protection. The first column is for two-photon NOON states, and the second is for anti-NOON states. The disordered region contains random on-site energies (i.e., the refractive indices) within the range [-*W*,*W*]. Each subfigure here shows the correlation map, $\Gamma_{mn}$, just for the edge waveguides. The initial state (shown at the bottom-left corner of each subfigure) is composed of two lobes, denoting the position of injection of photons. For the NOON states, these lobes lie along the diagonal (meaning that if one photon is observed centered on waveguide *n*, the other must be centered there as well). For the anti-NOON state, the opposite is true: if one photon is centered at n, the other must be centered on *n+l*, meaning the lobes lie across the diagonal from one another.

In Figs. 3(a) and 3(b) – which correspond to the non-disordered case – the NOON and anti-NOON states travel along the diagonal (corresponding to moving rightwards along the edge), and undergo some degree of diffractive broadening. The broadening is weak because the topological edge state has nearly linear dispersion (see Fig. 2(b)), although there is curvature at the top and bottom of the gap. The broadening can be reduced by decreasing the bandwidth of the input wavefunction, 2$E_b$ (resulting in a wider wavefunction). It can therefore be suppressed to any desired degree.

Figures 3(c) and 3(d) represent the case where the photons enter a disordered region (but are otherwise analogous to 3(a) and 3(b)). Note that for both the NOON and anti-NOON cases, the path-entangled photons pass through mostly unchanged. For the NOON state, the "NOONity" of the state is preserved. This is striking considering that photon correlations are highly sensitive to the phase of the one-particle propagator – therefore not only does topological protection preserves the one-particle state in the disordered region, but also the nature of the path entanglement of two-particle states. This happens despite the fact that the defects imbue the state with a random phase compared to the clean case. Animations of the dynamics of the correlation map shown in Fig. 3 are given in multimedia files of the EPAPS.

We also compare the protection of the two-particle state in the topological case to what occurs in the non-topological case. Perhaps the best way of doing this is to simply remove the bulk and study the edge in isolation: without the bulk to provide a buffer between the top and bottom edges of the lattice, the system ceases to be topological and backscattering is permitted. Therefore, we consider a 1D lattice with a coupling term of strength $t$ between sites (akin to Fig. 1). We launch two entangled photons in analogous wavefunctions into the 1D lattice, analogously to the top edge of the 2D topological case. Figure 4 shows the correlation map for the NOON state wavefunction (top row is clean case, bottom row is the disordered). The left column represents the initial state, and next columns represent the wavefunction after successively longer propagation distances. Clearly, the clean disordered cases behave entirely differently: as a result of backscattering, the NOONity of the photons is destroyed in the non-topological lattice. A similar picture emerges for the anti-NOON state.

To quantify this, we introduce a quantity that measures the nature of path entanglement: the "NOONity," $N$, of a two-photon state:

$$N \equiv \sum_{mn} \Gamma_{mm}\Gamma_{nn} - \Gamma_{mn}^2,$$

The larger this quantity, the more NOON-like is the state. It is zero for an unentangled state and negative for an anti-NOON state. In Fig. 5, we plot $N$ as a function of $z$, in order to compare the shape of the wavefunction with and without disorder. In 5(a) the NOON state is in the topologically trivial 1D array; 5(b) the anti-NOON state in the same; 5(c) the NOON state on the topological edge; and 5(d) the anti-NOON state in the same. In all cases in Fig. 5 (including the non-disordered case), there is some decrease in $N$ caused by unavoidable diffractive broadening (choosing a wavepacket covering ~10 waveguides leads to a decrease in NOONity of ~1/2 over 100 coupling lengths). However, in both 5(a) and 5(b), we see that the disorder has destroyed the NOON and anti-NOON states in the non-topological system. By comparison, in Fig. 5(c) and 5(d) – which show the topological case – we see that the character of the state is largely preserved. This is depicted both in Fig. 3 as well as in the multimedia files of the supplementary information section II. Clearly, the topological protection has protected the character of the state – preserving the "NOONity" of both the NOON and anti-NOON states. We account for the deviation between the clean and disordered topological cases as follows. When topological edge states propagate past a defect unscattered, they effectively "go around it," meaning it takes them an additional amount of time ($z$), compared with a clean system. We therefore interpret the difference between the clean and disordered case shown in Fig. 5(c) and (d) as the edge states taking "longer" to traverse the disordered path. As a result of diffractive broadening, there is some decrease in NOONity in the disordered case compared with the clean one (though much less than in the non-topological case).

In conclusion, we have shown that topological edge states can transport path-entangled multi-photon states in a robust way. This may lead to robust transport of quantum information through disordered environments, and provokes many new questions. For example, what happens to the topological protection in the presence of photon interactions? To what extent is photonic topological protection compatible with networks that are useful for quantum information?

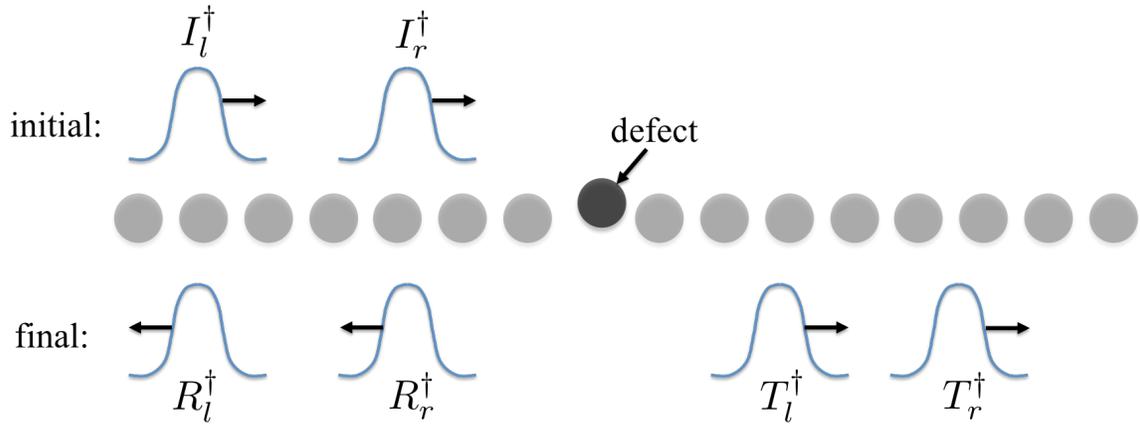

Figure 1: we consider an injected NOON state based on spatially separated input states. They scatter off a defect in the topologically trivial 1D lattice, resulting in reflected and transmitted wavepackets.

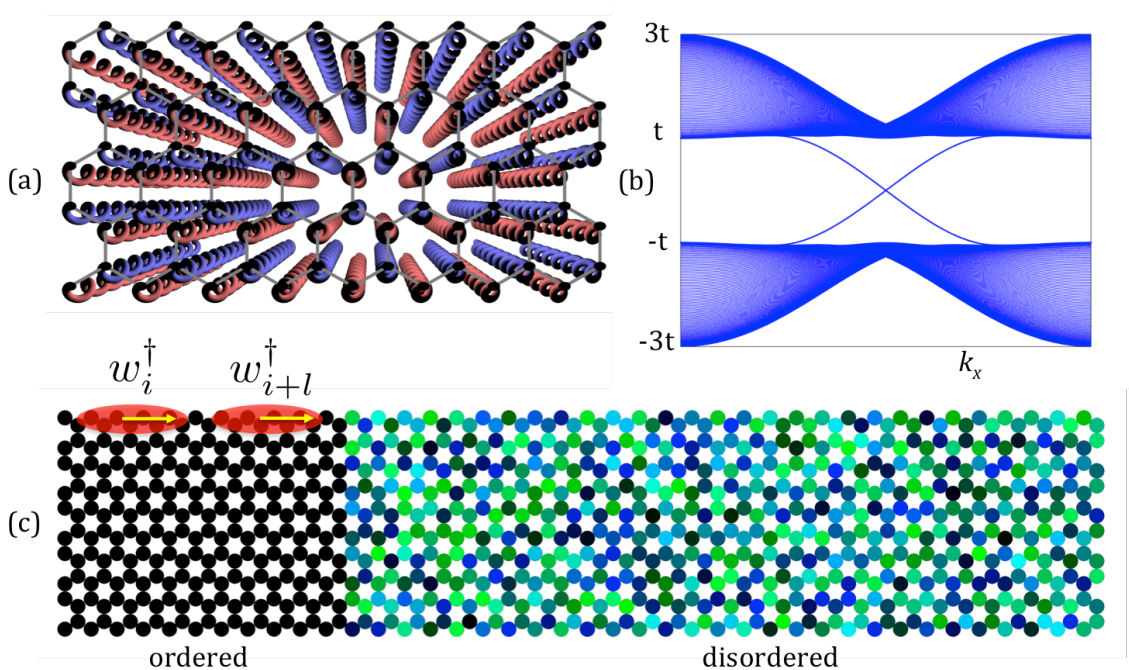

Figure 2. (a) Honeycomb lattice of helical waveguides forms a PFTI [9]. (b) Band structure in the topological case (edge states cross the band gap). (c) Probing effects of disorder: the two-photon state is injected in the 'clean' region (left), and enters the disordered region (right).

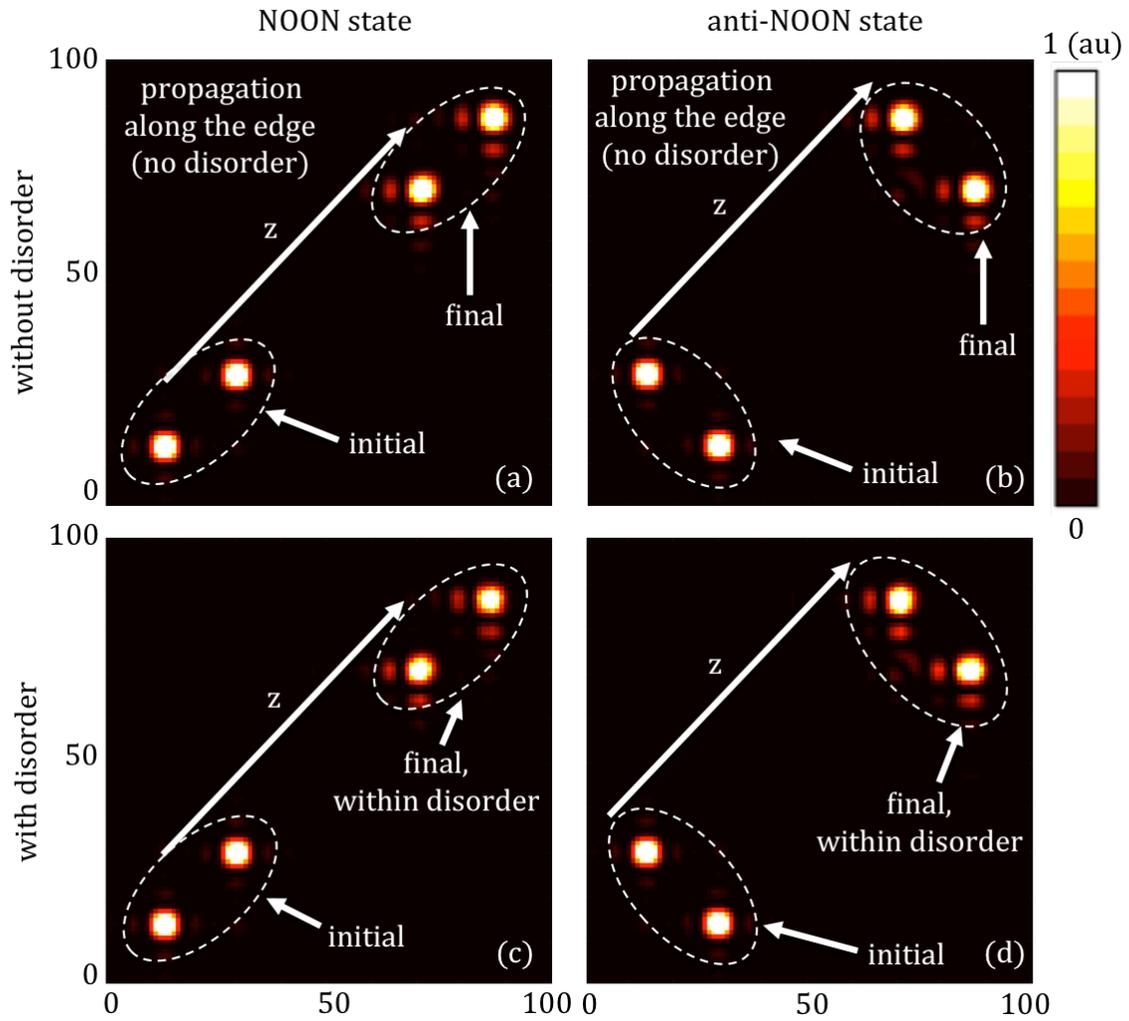

Figure 3. Correlation map evolution along the topological edge for: (a) NOON state with no disorder; (b) anti-NOON state with no disorder; (c) NOON state with disorder; (d) anti-NOON state with disorder. The disorder starts half way through. Here we see that the presence of disorder has not caused a strong change in the qualitative behavior of the correlation map (see Fig. 4 for a comparison with the topologically trivial case).

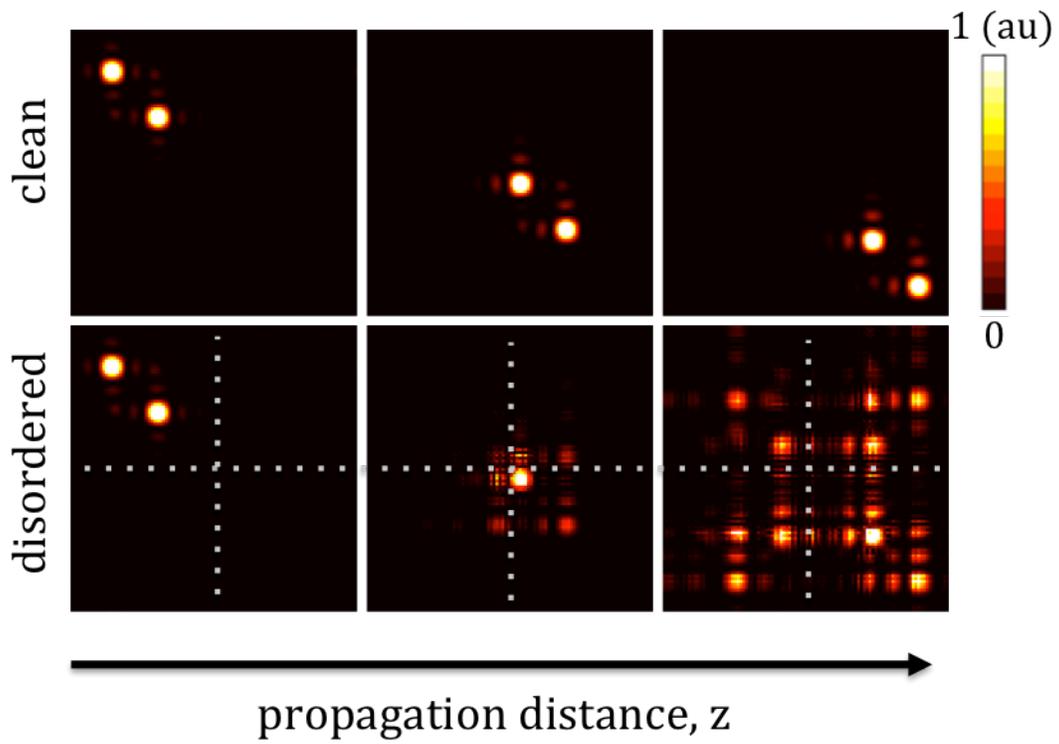

Figure 4. The figures show the correlation map evolution in the topologically trivial 1D array for NOON states, in two cases: without (top row) and with (bottom row) disorder present. It is clear that the defect destroys the nature of the photon correlations (disorder interface at dotted line).

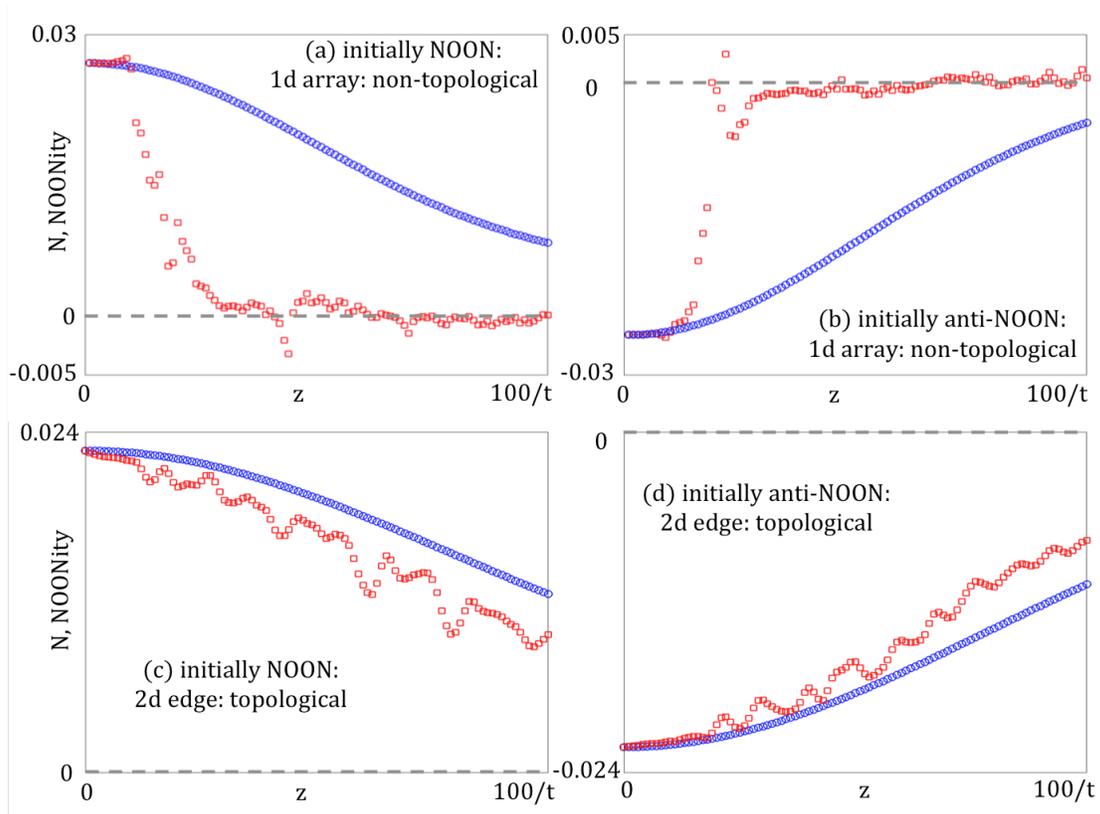

Figure 5. Evolution of the "NOONity" as a function of $z$ for: (a) a NOON state on the non-topological edge; (b) anti-NOON state on the non-topological edge; (c) NOON state on the topological edge; (d) anti-NOON state on the topological edge. In the topologically trivial cases (a) and (b), NOONity and anti-NOONity are destroyed by disorder, whereas in the topological cases (c) and (d) they are largely conserved. The blue points indicate the clean case (disorder, $W=0$); the red points indicate the disordered case ($W=t/2$).

Supplementary information: "Topological protection of photonic path entanglement"

In this section, we derive an expression for the photon correlation map $\Gamma$ given the initial two-photon wavefunction, namely:

$$\Gamma_{rs} = \left|\left(U\left(C_0 + C_0^T\right)U^T\right)_{rs}\right|^2, \quad (S1)$$

where $\Gamma_{rs}$ is 2-photon correlation map (at waveguides $r$ and $s$), $U$ is the single-photon propagator from time $0$ to $t$, and $C_0$ a matrix representing wavefunction at time $t = 0$.

We assume two photons injected into the system, and since photon number is conserved by our Hamiltonian, the wavefunction can always be written as:

$$\left|\psi(t)\right\rangle = \sum_{ij} C_{ij}(t) a_i^\dagger a_j^\dagger |0\rangle, \quad (S2)$$

where $a_n^\dagger$ is the creation operator for a photon in waveguide $n$. We denote $C_{ij}(t=0) = C_{0,ij}$. The hopping coefficient between two waveguides $n$ and $m$ is $t_{nm}$, and therefore our Hamiltonian can be written as:

$$H = \sum_{nm} t_{nm} a_n^\dagger a_m. \quad (S3)$$

We now aim to write the Schrödinger equation $i\partial_t |\psi\rangle = H|\psi\rangle$ as an equation for the coefficients $C_{ij}(t)$. We evaluate $H|\psi\rangle$:

$$H|\psi\rangle = \sum_{nm} t_{nm} a_n^\dagger a_m \sum_{ij} C_{ij}(t) a_i^\dagger a_j^\dagger |0\rangle =$$
$$\sum_{nmij} t_{nm} C_{ij}(t) a_n^\dagger a_m a_i^\dagger a_j^\dagger |0\rangle = \sum_{nij} t_{in}\left(C_{nj}(t) + C_{jn}(t)\right) a_i^\dagger a_j^\dagger |0\rangle. \quad (S4)$$

The last equality can be easily proven using the well-known commutation relations of the creation and annihilation operators. Using Eq. (S4), the Schrödinger equation is now written as:

$$i\partial_t \sum_{ij} C_{ij}(t) a_i^\dagger a_j^\dagger |0\rangle = \sum_{nij} t_{in}\left(C_{nj}(t) + C_{jn}(t)\right) a_i^\dagger a_j^\dagger |0\rangle. \quad (S5)$$

Using the following identity, again easily proven using commutation relations:

$$\langle 0| a_p a_q a_i^\dagger a_j^\dagger |0\rangle = \delta_{iq}\delta_{jp} + \delta_{ip}\delta_{jq}, \quad (S6)$$

we operate with $\langle 0|a_p a_q$ from the left on our Schrödinger equation (S5), and get for the left hand side of (S5):

$$i\partial_t \sum_{ij} C_{ij}(t) \langle 0|a_p a_q a_i^\dagger a_j^\dagger |0\rangle = i\partial_t \left(C_{qp} + C_{pq}\right) \qquad (S7)$$

For the right hand side of (S5) we get:

$$\sum_{nij} t_{in}\left(C_{nj}(t)+C_{jn}(t)\right)\langle 0|a_p a_q a_i^\dagger a_j^\dagger |0\rangle = \sum_n \left(t_{qn}\left(C_{np}(t)+C_{pn}(t)\right) + t_{pn}\left(C_{nq}(t)+C_{qn}(t)\right)\right) \qquad (S8)$$

We define a symmetric matrix $D$: $D_{mn} = (C_{mn} + C_{nm})/2$. Using the matrix $D$, our Schrödinger equation now takes the form:

$$i\partial_t D_{qp} = \sum_n \left(t_{qn} D_{np} + t_{pn} D_{nq}\right) \qquad (S9)$$

Eq (S9) can also be written in matrix form:

$$i\partial_t D = TD + (TD)^T \qquad (S10)$$

With $T_{mn} = t_{mn}$.

The solution to equation (S10) can easily be shown to be

$$D(t) = e^{-iTt} D(0) e^{-iT^T t}. \qquad (S11)$$

We write $U(t) = e^{-iTt}$, and recognize that the matrix $U(t)$ is the propagator of the Schrödinger for a single particle. We can assume, without loss of generality, that $C$ is also symmetric because only the symmetric part of $C$ evolves in time. Then, we have $D = C$ and we immediately get:

$$C(t) = U(t) C(0) U^T(t) \qquad (S12)$$

It is now possible to write the state at time in the following form:

$$|\psi(t)\rangle = \sum_{ij} C_{0,ij} \left(U^T a^\dagger\right)_i \left(U^T a^\dagger\right)_j |0\rangle \qquad (S13)$$

We now wish to evaluate the 2-photon correlation map if the state after propagating time $t$. That is, we wish to evaluate:

$$\Gamma_{rs} = \langle \psi(t) | a_r^+ a_s^+ a_s a_r | \psi(t) \rangle \qquad (S14)$$

We start by writing the bra and ket form of Eq (S13) in a more explicit manner:

$$|\psi(t)\rangle = \sum_{ijpq} C_{0,ij} \left( U^T_{ip} a_p^+ \right) \left( U^T_{jq} a_q^+ \right) |0\rangle \quad , \quad \langle \psi | = \langle 0 | \sum_{i'j'mn} C^*_{0,i'j'} \left( U^\dagger_{i'm} a_m \right) \left( U^\dagger_{j'n} a_n \right) \qquad (S15)$$

We now write the correlation at time t:

$$\Gamma_{rs} = \langle \psi(t) | a_r^+ a_s^+ a_s a_r | \psi(t) \rangle = \sum_{\substack{ijpq \\ i'j'mn}} C^*_{0,i'j'} U^\dagger_{i'm} U^\dagger_{j'n} C_{0,ij} U^T_{ip} U^T_{jq} \langle 0 | a_m a_n a_r^+ a_s^+ a_s a_r a_p^+ a_q^+ | 0 \rangle$$

We use the identity (again, easily proven using commutation relations):

$$\langle 0 | a_m a_n a_r^+ a_s^+ a_s a_r a_p^+ a_q^+ | 0 \rangle = \delta_{rm}\delta_{ns}\delta_{ps}\delta_{rq} + \delta_{rm}\delta_{ns}\delta_{pr}\delta_{sq} + \delta_{rn}\delta_{ms}\delta_{ps}\delta_{rq} + \delta_{rn}\delta_{ms}\delta_{pr}\delta_{sq} \qquad (S16)$$

And get the correlation:

$$\Gamma_{rs} = \sum_{\substack{ijpq \\ i'j'mn}} \langle 0 | C^*_{0,i'j'} U^\dagger_{i'm} U^\dagger_{j'n} \left( \delta_{rm}\delta_{ns}\delta_{ps}\delta_{rq} + \delta_{rm}\delta_{ns}\delta_{pr}\delta_{sq} + \delta_{rn}\delta_{ms}\delta_{ps}\delta_{rq} + \delta_{rn}\delta_{ms}\delta_{pr}\delta_{sq} \right) C_{0,ij} U^T_{ip} U^T_{jq} | 0 \rangle \qquad (S17)$$

By evaluating each term individually and regrouping the terms into matrix form, it can be shown about some algebra that:

$$\Gamma_{rs} = \left| \left( U (C_0 + C_0^T) U^T \right)_{rs} \right|^2. \qquad (S18)$$

Equation (S18) is the end result of this section.